\documentclass{phostproc}

\title{Current problems in stellar pulsation theory}
\author{M.-A. Dupret$^{1}$}

\affiliation{$^{1}$ STAR Institute, University of Liège, 19C Allée du 6 Août, B-4000 Liège, Belgium}

\shorttitle{Current problems in stellar pulsation theory}
\shortauthors{M.-A. Dupret}

\abs{The last decade lead to major progress in asteroseismology and stellar physics with the advent of space missions. Thanks to the richness and precision of current oscillation spectra, sophisticated seismic probing techniques allow us now to pinpoint the limits of our current models of stellar structure and evolution. However, the accuracy of the seismic diagnosis depends on the accuracy of the pulsation models. In solar-like oscillations, the main source of inaccuracy comes from the near-surface layers where the oscillations are non-adiabatic and strongly coupled with turbulent convection. Some pulsating stars rotate fast and this must be accurately taken into account in the modeling of their pulsations. In others, the magnetic field or the dynamic tides could play some role. I propose here an overview of the great achievements and current limitation of asteroseismology.}

\begin{document}

\maketitle

\section{Introduction}

Asteroseismology is now reaching its golden years thanks to the advent of space missions providing
us with wonderful lightcurves. Analysing them, the current generation of asteroseismologists
extracted oscillation spectra of unequaled richness. This already lead to major often unexpected
discoveries revolutionizing our view of stellar interiors. As a general introduction to the conference proceedings, I summarize here some of these achievements, identify the main limitations of current techniques and talk about future prospects.

\section{Great achievements of asteroseismology}

\subsection{Red giants}

The discovery of very rich spectra of solar-like oscillations in red giants and their interpretation is certainly one of the most important achievements of asteroseismology. It opened great and unexpected new horizons. First, precise measurements of masses, radii and ages of a huge amount of red giants observed by CoRoT and Kepler were possible. This was highly welcomed by galactic archaeologists, opening an entirely new interdisciplinary field. \citet{Miglio09} was the first to identify this connection. For more detail, I refer to its paper in these proceedings. One of the new step is now to combine asteroseismic measurements of mean densities with radii measurements with GAIA. I refer to the paper of Marc Pinsonneault (these proceedings) for more detail. Of course, the future is even brighter with PLATO \citep{Miglio17}. 

Second, the discovery of non-radial mixed modes in red giants appeared to be a golden gate revealing their hidden core. Rotational splittings of mixed modes were first discovered by \citet{Beck12}. They were used to measure the core rotation of numerous red giants \citep[see][for the last measurements]{Gehan18} with e.g. the method of \citet{Goupil13}. This revealed unexpected slow core rotation due to unknown braking processes \citep[see e.g.][]{Eggenberger12,Marques13,Eggenberger17}. Differential rotation could be measured in subgiants \citep{deheuvels14} and core helium burning stars \citep{deheuvels15}. The subgiants rotation rates seem in agreement with models of angular momentum transport by  plume-induced internal gravity waves \citep{Pincon17}. The period spacings of mixed modes allow us now to easily distinguish between hydrogen- and helium-burning red giants \citep{Bedding11}, to measure convective core overshooting and the masses of their helium cores \citep{Montalban13,Bossini15}.
Some of these great achievements are discussed in several papers of these proceedings by S. Deheuvels, D. Stello, M. Vrard and M. Takata. For a detailed state of the art of red giants asteroseismology as it was in 2017, see \citet{Hekker17}.

The future prospects for red giants’s seismology are great. Based on an improved asymptotic theory \citep{Takata16b}, stretching methods now enable us to transform the non-regular oscillation patterns of mixed modes into regular ones \citep[see][for the most recent update]{Mosser18}. Three families of seismic indicators arise: the classical ones associated to the acoustic cavity (like in main-sequence solar-like stars), those related to the gravity modes cavity: period spacing, gravity offset, core rotation, splittings and, icing on the cake, the coupling factor probing the evanescent zone. We can already learn a lot about stellar interiors with that ! Beyond that, buoyancy glitches can be looked for and interpreted \citep{Cunha15}; and inversion methods able to include mixed-modes could be developed.

The future prospects for subgiants and young red giants are also great. With the method developed by \citet{deheuvels17}, it is now possible to probe the core rotation of red giants from their asymmetric splittings. The pending question is how to deal with the non-linearity associated to the avoided crossings with mixed modes? Both forward modeling and inversion techniques must be adapted to this reality. And this is not a simple problem, because contrary to more evolved red giants, modes are out of the asymptotic regime in the g-cavity, so that the stretching is not easy.

\subsection{g-modes}

The discovery of dense spectra of g-modes in different types of stars is a second great achievement. Series of consecutive modes are now identified in $\gamma$~Dor \citep[see e.g. ][]{vanreeth15b} and SPB stars \citep[see e.g.][]{Papics17}, allowing us to measure accurately their core rotation and the extension of their convective core \citep{ouazzani17,christophe18,Vanreeth16}. More detail about that is given by Bedding and Ouazzani (these proceedings). Dense spectra of g-modes are also observed in extreme horizontal branch stars. Although the interpretation of these spectra is a matter of debate \citep{Reed11,Ostensen14,charpinet14}, this holds the hope of a very detailed seismic probing of these stars in the near future (see Charpinet, these proceedings). 
Detailed seismic probing of the internal composition of white dwarfs is now possible as shown recently by \citet{Giam18}, see also the review of Hermes (these proceedings). Finally, there is also the possible discovery of high-order g-modes in the Sun \citep[see the paper in these proceedings]{Fossat17,Fossat18}, although some consider it as ``fragile'' \citep{Schunker18}.

\subsection{p-modes}

Very rich oscillation spectra of p-modes are now detected in solar-type stars. Of particularly high quality is the so-called  Kepler LEGACY Sample of stars, for which accurate seismic measurements of radii, masses and ages are possible \citep[see e.g.][]{silva17}. The precision of the frequencies and the number of detected modes makes it possible to extend seismic inversion techniques initially developed for the Sun to these stars \citep{Buldgen15b}. Internal mixing \citep{Buldgen15a} and core overshooting \citep{deheuvels16}  can now be seismically measured. Internal rotation could be probed, revealing nearly uniform internal rotation along the radial axis  \citep{Benomar15} and latitudinal differential rotation, the equator rotating approximately twice as fast as their midlatitudes \citep[][and these proceedings]{Benomar18}.  

New methods are being developed for the seismic probing of solar-like stars. A first path is to go beyond the usual seismic indicators and introduce new ones. One promising path is the phase matching method proposed by \citet{Roxburgh16}. Another one was recently proposed by \citet{farnir} (and these proceedings). A lot of work is also done on the development of new kinds of optimization algorithms for forward seismic modeling. For the future, the development of non-linear inversion methods could also be envisioned.

\subsection{Rossby modes}

The discovery of rotation related modes is also an important recent gift. Particularly interesting are the global Rossby modes which appear to be detected in many types of stars: $\gamma$~Dor stars \citep{Vanreeth16}, spotted A and B stars, bursting Be stars and the heart beat stars \citep{saio18a}. The latter are particularly interesting for the study of the coupling between oscillations and tidal forces, as detailed in Guo (these proceedings). More detail on the high potential of these modes for seismic probing and their interesting properties is given in \citet{saio18a} and Saio (these proceedings).


\section{Current limitations of asteroseismology}

However, we must not forget the limitations of present asteroseismic techniques. The main current approach is forward modeling, but inversion begins to be also possible. The main specific limitation of the forward modeling approach is that it reduces the richness and complexity of stellar evolution to a small number of parameters to be determined. On the one hand, there are the physical parameters: mass, age, X, Z and, if included in the models, the rotation rate $\Omega$. On the other hand, there are parameters such as the mixing-length parameter $\alpha$, the overshooting parameter $\alpha_{ov}$, turbulent diffusion coefficients, … These last parameters are associated to very approximate models of convection (typically the MLT) and chemical transport. The results obtained by this approach are thus intrinsically limited. They are also model dependent since they depend on the choice of the opacity table, the equation of state, the convection treatment (MLT versus FST, instantaneous versus diffusive overshooting, \ldots), the initial chemical mixture,~\ldots A first limitation of inversion techniques is that they are linear, which requires a good reference model \citep[see][for the inaccuracies introduced by non-linearity in seismic inversions]{Buldgen17d} and complicates their application to stars with mixed modes. Their second limitation is that they require many identified modes, which is currently only the case with the longest Kepler lightcurves of solar-like stars. However, \citet{Buldgen19} showed recently that seismic inversion of the mean density of red giants is also possible, based on their radial modes only. Common limitations of forward modeling and seismic inversion are the surface effects problem and the standard approximations neglecting fast rotation, strong magnetic field, tidal effects and non-linearity in oscillation models.


\subsection{Model dependence}

It is useful to consider with a little more attention the problem of the model dependence associated with the small number of parameters defining standard models. This problem is coupled with the small number of available independent seismic indicators in many cases such as in ensemble asteroseismology of red giants. Fitting a red giant with two parameters (its age and mass) is less than the 4 parameters of the von Neumann’s elephant! \ldots It should also not be forgot that asteroseimology probes the interior of a star as it is now, not its evolutions and its associated long time-scale processes such as atomic diffusion, nuclear burning and macroscopic transport processes. Many papers in these proceedings are devoted to stellar physics. I just summarize here the main sources of uncertainty. 

Concerning first the microphysics, we have the ubiquitous atomic diffusion. Neglecting it in forward seismic modeling introduces systematic inaccuracies in e.g. age measurements. It is still treated approximately in most stellar evolution codes: partial ionization is generally neglected, metals are treated as a whole and radiative levitation is neglected. Some stellar evolution codes treat the diffusion element by element, include radiative forces and couple microscopic transport with macroscropic mixing (turbulence, thermoaline convection, \ldots). This is important but the cost in term of computation time is huge. I refer to Deal (these proceedings) for more detail.

Opacity computations are still approximate. Indeed, it is not possible yet to include all electron transitions and take into account the coupling between all states into account; a compromise is unavoidable. Opacities directly affect the temperature gradient and thus oscillation frequencies and ages \citep[see e.g.][fig.~18]{Lebreton14b}. 
The new abundance determinations by \citet{Asplund09} lead to significant discrepancy with the seismically inverted sound speed profile, the so-called solar problem. A local increase of the opacity just below the convective envelope is the most probable path to solve this problem \citep{Basu08}. However, the recent new opacity computations by the Los Alamos National Laboratory \citep{Colgan16} and by the CEA \citep{Mondet15} did not allow to solve the problem \citep{Buldgen17c}. This problem also illustrates the impact of the chemical mixture on the opacities. Erroneous assumptions on the chemical mixture and, in particular, assuming homogeneity of metal abundances is a source of errors. So, it is clear that opacities are still a source of unknown systematic inaccuracy in forward seismic modeling, either due to intrinsic inaccuracies in present opacity computations or due to inaccurate chemical mixture's assumptions. We must not forget also the uncertainties related to the equation of state. They are particularly important in brown dwarfs and probably also in white dwarfs. For more detail, I refer to Pain (these proceedings).

Macroscopic processes are subject to even larger uncertainties. The so-called rotational mixing hides in reality a complex interplay between angular momentum transport, chemical mixing, magnetism, tidal effects, mass loss, ... A state of the art of the problems associated to the modeling of these processes can be found in e.g Buldgen et al. (these proceedings) and their impact is discussed in e.g. \citet{Meynet16}. The transport of angular momentum by waves and modes is still very difficult to quantify. However, a new model of waves generation by penetrative convection recently proposed by \citet{Pincon16} could explain the internal rotation of subgiants \citep{Pincon17}.

Finally, there is of course the complexity of convection: on the one hand the uncertainties related to overshooting and semi-convection above convective cores (see Buldgen, these proceedings, for more detail), and on the other hand the uncertainties associated to convective envelopes and their coupling with oscillations, which I discuss in the next section.

\subsection{Surface effects}

This leads me to consider the so-called surface effects problem. This warrants indeed a special attention. What are surface effects ? In a nutshell, inaccurate modeling of the superficial layers affects the frequencies of high-order p-modes and leads thus to inaccurate seismic inferences. It must not be forgot that there are two sources of inaccuracies. On the one hand, the structural inaccuracies mainly associated to the modeling of convection in atmosphere models, and on the other hand the modal inaccuracies associated to the adiabatic approximation (neglecting thus the fact that oscillations are nonadiabatic and the coupling between oscillations and convection is strong in superficial layers).



\subsubsection{Structural inaccuracies}

As detailed in Ludwig (these proceedings), 3D atmosphere models are now on the market. But how to use them appropriately for stellar evolution and asteroseismology is still under development. A first approach, the simplest one, is to use them to calibrate empirical frequencies corrections. The most recent work in this direction was done by \citet{Sonoi15}, \citet{Ball16} and \citet{Trampedach17}. A second approach is to use the 3D atmospheres to calibrate the convection parameters of the approximate convection models used in our stellar evolution codes. Most recent work in this direction was done by \citet{Trampedach14a}, \citet{Magic15}, \citet{Sonoi18} and these proceedings. Finally, interpolation in 3D grids can also be envisioned. Preliminary work in this direction was recently done by \citet{Jorgensen18b}.

\subsubsection{Modal inaccuracies}
Modal inaccuracies are another piece of cake\ldots  Oscillations are totally non-adiabatic near the surface. Moreover, the convective, thermal and oscillation time-scales are of the same order in the outermost layers of solar-like oscillators. Time-Dependent Convection (TDC) models are thus needed. I worked on that and I am strongly convinced that current models are by far too approximate and in many cases do not even catch the real physics of the coherent interaction between convection and oscillations. A few linear non-adiabatic oscillation models of the time-dependent interaction between convection and oscillations have been proposed and implemented. First, there is the model of \citet{Balmforth92t}, which is a non-local generalization of the MLT theory of \citet{Gough77}, widely used by G.~Houdek and his collaborators. Second, there is the model of \citet{Gabriel96} and  \citet{Grigahcene05}, which is based on the approach originally proposed by \citet{Unno67}. These two MLT perturbative theories are compared in \citet{Houdek15}. Finally, there is the even more complex TDC model developed by \citet{Xiong15}. All these models are clearly reaching their limits: on the one hand they encounter difficulties to fit observations (typically the mode line-widths, see below), and on the other hand, their complexity hides crude approximations. It is time to start trying to model this problem in all its 4D (3D space + time) complexity if we want to go out of this deadlock.

The good point which can help to progress at this level is that there are additional seismic constraints associated to solar-like stochastic excited oscillations: on the one hand the linewidths in the power spectrum, which are directly related to the mode damping rates, and on the other hand the amplitudes. The theoretical damping rates are obtained with non-adiabatic oscillation models including time-dependent convection and the theoretical amplitudes require the use of a stochastic excitation models, too. Confrontation to the observed values constrains thus these models and gives their more weight when they are used to model surface effects. The most recent confrontations with mode linewidths of Kepler stars are presented in Houdek (these proceedings) and \citet{Aarslev18}.

\subsubsection{Non-adiabaticity in classical pulsators}

In other types of pulsating stars, non-adiabaticity has a negligible impact on the frequencies, so no problem of surface effect for them, which is a big advantage. But that does not mean that nonadiabatic modeling is useless for these stars. It enables to understand and characterize the driving processes at the origin of pulsations. Moreover, the predicted range of excited modes, amplitude ratios and phases can be computed and compared with observations. This provides strong constraints on the opacity in $\beta$~Cep and SPBs \citep[e.g.][]{Walczak13,Salmon12, Dasz05,Dupret04}, in sdBs and in hot white dwarfs \citep{Quirion09}. Since the opacities depend on the chemical composition, constraints on chemical transport processes can also be obtained, as shown in \citet{Hu11}. In colder stars, this provides tests of time-dependent convection models and their current limitations \citep{Dupret05ds,Dupret05gd}. 

\subsection{Fast rotation}

The last limitation of oscillation models I consider here is the usual separation in spherical harmonics. My focus is on the effect of fast rotation. Fast rotation breaks the spherical symmetry and transforms the usual 1D eigenvalue problem into a 2D non-separable one. Codes solving rigorously the pulsation equations in this 2D framework have been implemented \citep{ouazzani12,Reese06}. They provide an entirely new view of fast rotating stars' pulsations, but the price in computation time is huge, making usual seismic probing methods impractical.

However, in gravito-inertial modes and global Rossby modes, the separation of variables obtained within the so-called traditional approximation appears to be a good compromise between fast computation and accuracy \citep{Ballot12}. In particular, using it appears to be justified for the interpretation of the wonderful oscillation spectra detected in Kepler gamma Dor stars \citep{ouazzani17}. The asymptotic theory within the traditional approximation can be used to disentangle their oscillation spectra \citep{christophe18,Bouabid13}. It shows that the two main seismic quantities that can be measured are the buoyancy radius $\Pi=(\int N/r \,dr)^{-1}$, directly related to the size of their convective core and their average core rotation $\int \Omega N/r\, dr\,\Pi$. Differential rotation could also be detected \citep{Vanreeth18}. The question for the future is: can we get more than these two measurements? I think the answer is yes. First, it is well known that trapping is possible in the $\mu$-gradient region, leading to oscillations of the period spacing \citep{Miglio08}. It could be used to constrain the sharpness of the chemical transition . Second, when looking more closely to observations and theoretical predictions, dips are sometimes present in the period spacing, which seem to be associated to differential rotation and/or mode coupling not taken into account by the traditional approximation \citep[][and Ouazzani, these proceedings]{saio18b}.

There are however cases where the variable separation is not justified: the fast rotating $\delta$~Sct and Be stars are the clearest example. In $\delta$~Sct stars, the equivalent of the large separation can be detected and used to measure their mean density \citep{GH09}. 
\citet{Mirouh19} (and these proceedings) developed a very promizing method of mode classification in these stars based on neural network, which could help for mode identification. Important theoretical work was also achieved for mode identification based on other observables, but this remains very difficult \citep{Reese17}. Another major difficulty remains the huge challenge of computing realistic evolutionary models of fast rotating stars near their break-up velocity. A lot of work has been done at this level, e.g. in the frame of the ESTER project \citep{Rieutord13}, but the problem is far from being fully solved. 

\subsection{Magnetic field, tidal effects, non-linearity}

Current pulsation models usually neglect magnetic and tidal effects. Fortunately, this is mostly justified. For the magnetic field, the only major exception is the modeling of ro~Ap pulsations, in which the Lorentz force has a significant  dynamical effect on pulsations. A non-perturbative model for axisymmetric p-mode pulsations of stars with dipole magnetic fields was developed by e.g. \citet{Saio04}. More recently, \citet{Loi} (and these proceedings) analysed in detail the effects of a strong magnetic field on internal gravity waves, an analysis which can find application in various astrophysical contexts, including the dipole dichotomy problem in red giants, the solar interior, and compact star oscillations. Studying the impact of tidal forces on pulsations in close binaries is still in its infancy. This problem recently got new attention with the detection of tidally excited oscillations in heartbeat stars observed by Kepler \citep{Guo17} (and these proceedings), with first models developed by \citet{Fuller17}. The linear approximation is ubiquitous in oscillation models used in asteroseismology. Currently, it seems unavoidable. On the opposite, non-linear pulsation models are widely used for the modeling of high amplitude radial pulsations and could help to explain longstanding problems such as the Blazhko effect (Koll\'ath, these proceedings).





\bibliographystyle{phostproc}
\bibliography{dupret-ref.bib}

\end{document}